\documentclass{ws-mpla}
\usepackage{lineno}
\usepackage{wrapfig}
\begin{document}
\markboth{Lokesh Kumar}
{Review of Recent Results from the RHIC Beam Energy Scan}

\catchline{}{}{}{}{}

\title{REVIEW OF RECENT RESULTS FROM THE RHIC BEAM ENERGY SCAN
}

\author{LOKESH KUMAR}

\address{
 School of Physical Sciences, National Institute of Science
 Education and Research, Bhubaneswar, Odisha 751005, India\\
~~\\
lokesh@rcf.rhic.bnl.gov; lokesh@niser.ac.in}


\maketitle


\begin{abstract}
We review recent results from the RHIC beam energy scan (BES) program,
aimed to study the Quantum Chromodynamics (QCD) phase diagram. 
The main goals are to search for the possible phase boundary,
softening of equation of state or first order phase transition, and
possible critical point. Phase-I of the BES program has recently concluded
with data collection for Au+Au collisions  at center-of-mass energies
($\sqrt{s_{NN}}$) of  7.7, 11.5, 19.6, 27, and 39 GeV. 
Several interesting results are observed for these lower energies
where the net-baryon density is high at the mid-rapidity. These
results indicate that the matter formed at lower energies (7.7 and
11.5 GeV) is hadron
dominated and might not have undergone a phase transition. In addition, the
centrality dependence of freeze-out parameters is observed for the
first time at lower energies,
slope of directed flow for (net)-protons measured versus rapidity shows an interesting behavior at lower energies, and higher moments of net-proton show deviation from Skellam expectations at lower energies.
An outlook for the future BES Phase-II program is presented
and efforts for the detailed study of QCD phase diagram are discussed.

\end{abstract}

\ccode{PACS Nos.: 25.75.-q,25.75.Nq, 12.38.Mh,25.75.Dw,25.75.Gz,25.75.Ld}
\section{Introduction}	
\begin{figure}[htbp]
\begin{center}
\includegraphics[width=0.55\textwidth]{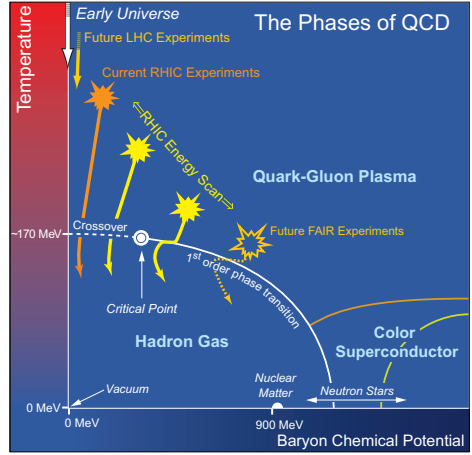}
\end{center}
\caption{(Color online) 
Schematic QCD phase diagram plotted as temperature $T$ versus baryon
chemical potential $\mu_{B}$.
}
\label{fig_bes}
\end{figure}
The main goals of high-energy heavy-ion collision experiments are to
search and 
study the
hot and dense matter called Quark-Gluon Plasma (QGP) formed in these
collisions~\cite{qgp}. Moreover, there is a great interest in understanding the
QCD phase diagram, a phase diagram for strong interactions, to the
level of that for electromagnetic interactions such as water. 
The results from top RHIC energies suggest the formation of
QGP~\cite{qgp}. 
The focus has now shifted to study the QGP properties~\cite{qgp_prop} and establish the QCD phase
diagram. In this review, we will concentrate on the latter part
which is establishing the QCD phase diagram through a dedicated
program at RHIC called beam energy scan program~\cite{ref_bes,kumar_npa,mohanty_npa}.
Figure~\ref{fig_bes} shows the schematic QCD phase diagram plotted as
temperature $T$ vs. baryonic chemical potential $\mu_B$~\cite{nsac}. There are two
main phases predicted in the QCD phase diagram: QGP and hadronic gas phase.
Lattice QCD calculations predict that the transition between QGP and
the hadronic gas at baryonic chemical potential $\mu_{B}=$ 0 is a crossover~\cite{lattice1}. At large $\mu_{B}$, 
the transition between QGP and hadron gas is expected to be a first order phase
transition~\cite{ejiri,kapusta}. Subsequently, the end-point of this
first-order phase transition line (while going towards the crossover)
would be the position of a critical point~\cite{cp}.
While there is a little guidance from the theory side about the QCD
phase diagram, efforts are ongoing from the experimental side to
establish some of its distinct structures such as phase boundary between
de-confined phase of quarks and gluons and hadron gas phase,
first-order phase transition line, and the critical point.

Experimentally, the two axes: $T$ and $\mu_B$ of the QCD phase diagram can be obtained from the momentum distributions and the ratios of the produced particles in heavy-ion collisions. Each collision energy corresponds to one $T$-$\mu_B$ point in the phase diagram. 
So, idea is to collect
data at different center-of-mass energies by colliding heavy-ions.
Once, the $T$-$\mu_B$ point is obtained, one can
look at the various signatures for the phase boundary, first-order
phase transition, and
the critical point. One of the interesting aspect is to locate the
energy where the established signatures of the QGP (at top RHIC energy)
disappear or ``turn-off''. This is how the RHIC beam energy scan
program was planned~\cite{ref_bes,kumar_npa,mohanty_npa}.
The proposal for the BES program was made in the year
2008.~This was followed by a successful data taking and physics
analysis of a test run of Au+Au collisions below injection energies at $\sqrt{s_{NN}}$ = 9.2 GeV~\cite{9gev}.
The first phase of the BES program was started in
the year 2010 with data taking in Au+Au collisions at three low energies of 7.7, 11.5, and
39 GeV. In 2011, two more energies at $\sqrt{s_{NN}}=$ 19.6
and 27 GeV were included.
Table~\ref{ta1}
lists various energies and corresponding number of events 
collected by the STAR detector in 2010--2011 for
Phase-I of the BES program.
\begin{table}[h]
\tbl{The data collected during the Phase-I of the BES program.}
{\begin{tabular}{c|c|c}
\hline
Year & $\sqrt{s_{NN}}$ (GeV) & $N_{\rm{event}}$ (Million)  \\
\hline
2010   & 7.7   & 5       \\
2010  & 11.5   &  12\\ 
2010  & 39   & 130     \\
2011  &  19.6   & 36      \\
2011  & 27   & 70      \\
\hline
\end{tabular}\label{ta1} }
\end{table}

This review is organized as follows. 
In Sec. 2, we discuss freeze-out parameters
that provide information about $T$-$\mu_B$ points in the QCD phase
diagram. In Sec. 3, signatures of first-order phase transition and
that for “turn-off” of QGP are dis- cussed. These include results on
freeze-out eccentricity, directed flow, elliptic flow, dynamical
charge correlations, and nuclear modification factor. The signatures
for the search of possible critical point are discussed in Sec. 4 that
include energy dependence of particle ratio fluctuations and higher
moments of conserved quantities such as net-proton. Section 5 provides
the outlook for the BES Phase-II program. Finally, we conclude with a
summary in Sec. 6.

\section{Freeze-out Parameters}
\begin{figure}[htbp]
\begin{center}
\includegraphics[width=0.45\textwidth]{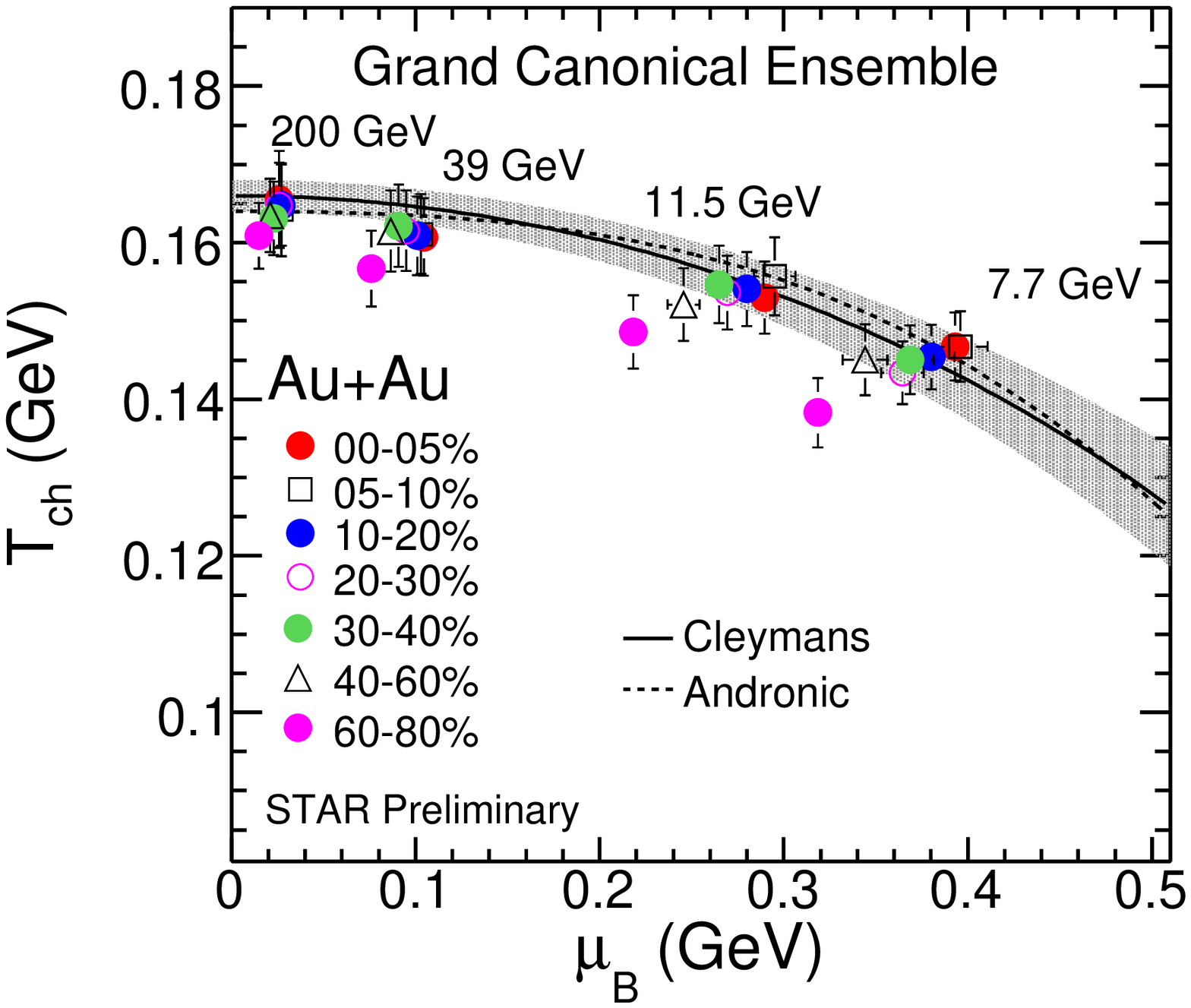}
\includegraphics[width=0.45\textwidth]{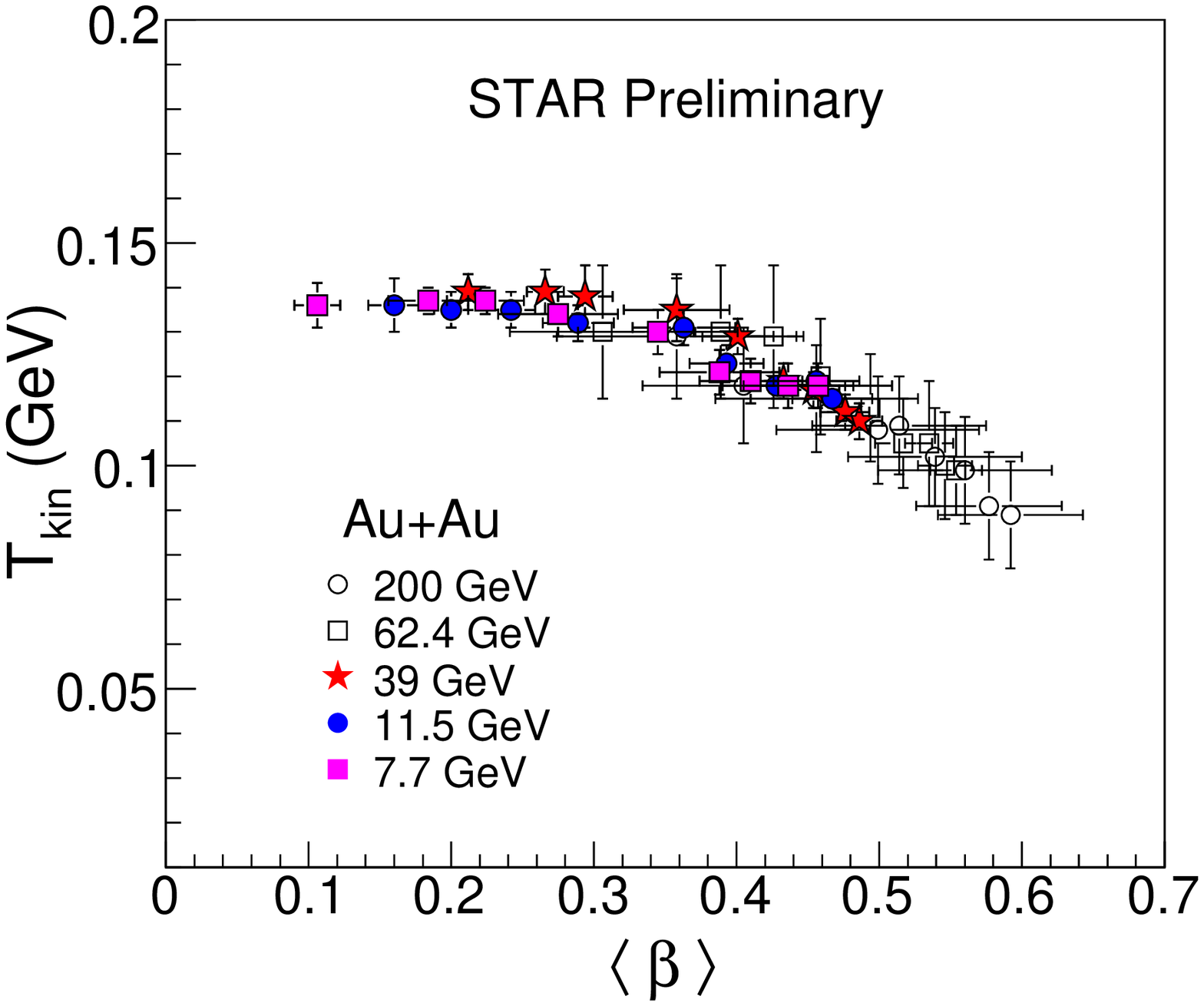}
\end{center}
\caption{(Color online) 
Left panel: Variation of $T_{\rm{ch}}$ with $\mu_{B}$ for
different energies and centralities. 
Right panel: Variation of
$T_{\rm{kin}}$ with $\langle \beta \rangle$ for different energies
and centralities. Errors in both panels represent the quadrature
sum of systematic and statistical errors.}
\label{chem_kin}
\end{figure}

The QCD phase diagram is the variation of temperature $T$ and baryon
chemical potential $\mu_B$.
These quantities can be 
extracted from the measured hadron yields.
Transverse momentum spectra for the BES Phase-I energies are
obtained for $\pi$, $K$, $p$, $\Lambda$, $\Xi$, $K^{0}_{S}$, and
$\phi$~\cite{lok,zhu}. From these distributions, corresponding particle yields are
obtained and various particle ratios are constructed.
These particle ratios are used to obtain the chemical freeze-out (a
state after the collision
when the yields of particles get fixed)
conditions using the statistical thermal model
(THERMUS)~\cite{stm,Andronic:2008ev,Wheaton:2004qb}. The two main extracted parameters are chemical
freeze-out temperature $T_{\rm{ch}}$ and $\mu_{B}$. 
Figure~\ref{chem_kin} (left panel) shows the variation of the extracted
chemical freeze-out parameters using the Grand-Canonical Ensemble (GCE) approach of
THERMUS for different energies and
centralities~\cite{Kumar:2012fb,Das:2012yq}. The curves represent the
parameterizations of $T_{\rm{ch}}$ and $\mu_B$~\cite{Andronic:2009jd,Cleymans:2005xv}.
We observe that at top RHIC energy, there is a little variation of
chemical freeze-out parameters with centrality. While at
lower energies, $T_{\rm{ch}}$ shows a variation with $\mu_{B}$ as a
function of centrality. The centrality dependence of these parameters
is observed for the first time in heavy-ion collisions at these lower
energies. One advantage of having such a dependence is that one can
explore larger portion of the QCD phase diagram.

The particle spectra can be used to obtain the kinetic freeze-out
(a state after the collision
when the spectral shapes of particles get fixed)
conditions using the Blast Wave (BW) model~\cite{Schnedermann:1993ws}. The BW model
is used to simultaneously fit the $\pi$, $K$, $p$ spectra and the two
main extracted parameters are kinetic freeze-out temperature $T_{\rm{kin}}$
and average flow velocity $\langle \beta
\rangle$. Figure~\ref{chem_kin} (right panel) shows the variation of
kinetic freeze-out parameters for different energies and
centralities~\cite{Das:2012yq}. 
We observe that at a given collision energy, there is an anti-correlation
between $T_{\rm{kin}}$ and $\langle \beta \rangle$. For a given
collision centrality, the freeze-out temperature at high energy is lower and the
average collectivity velocity $\langle \beta \rangle$ is larger due to expansion. 

\section{Search for First Order Phase Transition \& Turn-off of QGP
  Signatures}
Having discussed about accessing the QCD phase diagram by obtaining
$T-\mu_B$ points, 
we can now discuss various signatures for first order phase
transition or softest point in equation of state and those showing ``turn-off'' of QGP. These include
freeze-out eccentricity, directed flow, elliptic flow, dynamical
charge correlations, and nuclear modification factor.

\subsection{Freeze-out Eccentricity}
Eccentricity at freeze-out can be extracted as:
$\epsilon_F=\frac{\sigma_y^2 - \sigma_x^2}{\sigma_y^2 + \sigma_x^2}
\approx 2 R_{s,2}^2/R_{s,0}^2$,
where $\sigma_x$ and $\sigma_y$ correspond to the widths of the
participant zone at freeze-out in the in-plane and and out-of-plane
directions, respectively~\cite{Mount:2010ey}. $R_{s,2}^2$ and $R_{s,0}^2$ are the 2$^{nd}$-order and
0$^{th}$-order Fourier coefficients radius terms along the ``side'' direction
(perpendicular to the direction of average transverse pair momentum or ``out''
and that along the beam direction or ``long''), respectively. The ratio
$R_{s,2}^2/R_{s,0}^2$ is less affected by flow so it carries mainly
the geometric information.
\begin{figure}[htbp]
\begin{center}
\includegraphics[width=9.cm,height=6cm]{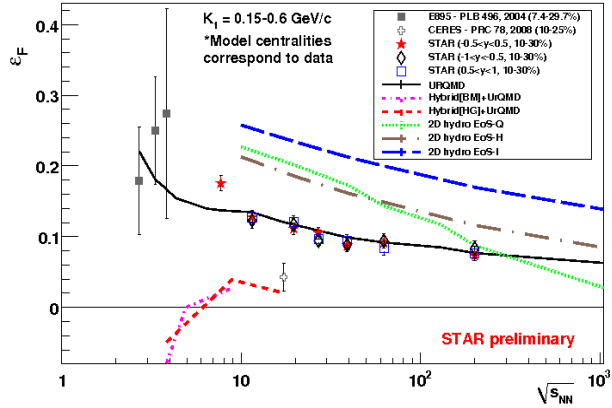}
\end{center}
\caption{(Color online) 
Freeze-out eccentricity as a function of beam energy compared to
different model calculations. }
\label{fig_hbt}
\end{figure}
 Freeze-out eccentricity may provide important
information related to both the equation of state and dynamical
processes involved in heavy-ion collisions as explained below. 
In non-central collisions, there is an initial anisotropy created
(elliptic shape) that
leads to more
compression along the shorter axis and hence larger initial pressure
gradients. This might lead to expansion along the shorter axis thereby
reducing the eccentricity. Ultimately, the system must evolve to a more round
freeze-out shape. Increasing energy would lead to longer lifetimes and
pressure gradients, and hence a monotonically decreasing excitation
function for the freeze-out eccentricity would be expected. 
If the system undergoes a first-order phase transition, a mixed phase
is expected. The system could spend more time in the mixed phase
compared to that in other phases. This may lead to different
expansions for different phases and hence non-monotonic freeze-out
shape. Figure~\ref{fig_hbt} shows the energy dependence of freeze-out
eccentricity compared to several model calculations including UrQMD~\cite{Bleicher:1999xi}
and other 2D hydrodynamical models~\cite{Lisa:2011na}. 
These results suggest a monotonic decrease in the freeze-out eccentricity with beam energy.

\subsection{Directed Flow}
The directed flow $v_1$ is calculated as $\langle \cos(\phi-\Psi_1)
\rangle$, where $\phi$ and $\Psi_1$ are 
the azimuthal angle of the produced particles and orientation of the
first-order event plane, respectively. 
The directed flow measurements near midrapidity for protons are
proposed to be sensitive to the equation of state
(EOS)~\cite{Brachmann:1999xt,Csernai:1999nf,Stoecker:2004qu}. It has been predicted that proton $v_1$ slope
show a non-monotonic behavior as a function of beam energy
illustrating change of
sign from positive to negative at lower energies and again going back
to positive at higher energies~\cite{Stoecker:2004qu}. This is sometimes called collapse
of proton flow. The minimum in proton $v_1$ slope is proposed to
correspond to a softest point in equation of state.
\begin{figure}[htbp]
\begin{minipage}{0.45\linewidth}
\centering
\vspace{-0.38cm}
\includegraphics[width=0.9\textwidth]{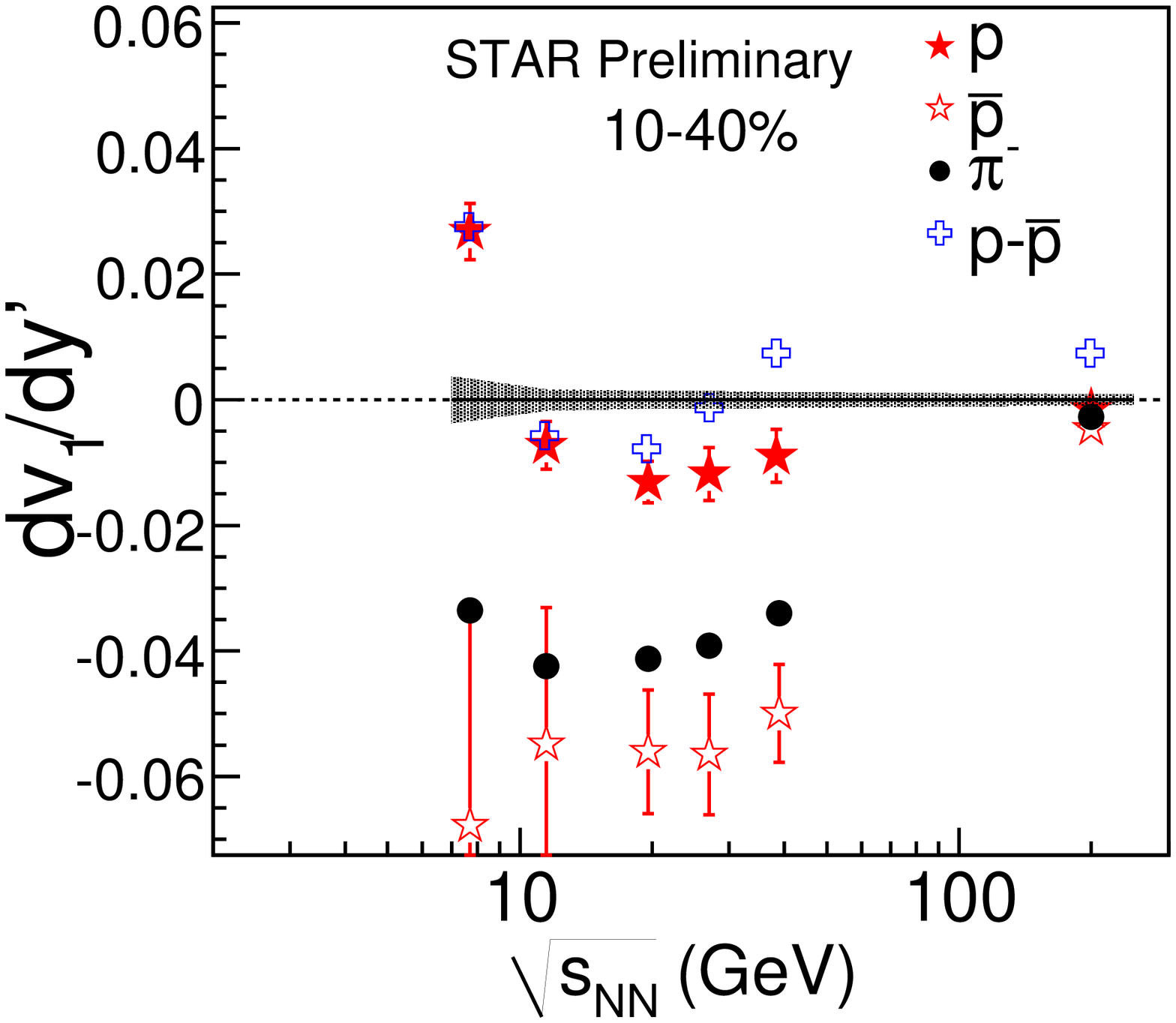}
\vspace{-0.5cm}
\caption{(Color online)
Directed flow slope
($dv_1/dy'$, $y'$=$y/y_{\rm{beam}}$) 
for $\pi^{-}$, $p$, $\bar{p}$, and net-protons ($p$-$\bar{p}$) near midrapidity as
a function of beam energy for mid-central (10--40\%) Au+Au
collisions. 
The shaded band refers to the systematic uncertainty on
net-proton measurements.
}
\label{fig_v1}
\end{minipage}
\hspace{0.4cm}
\begin{minipage}{0.45\linewidth}
\centering
\vspace{-0.38cm}
\includegraphics[width=0.9\textwidth]{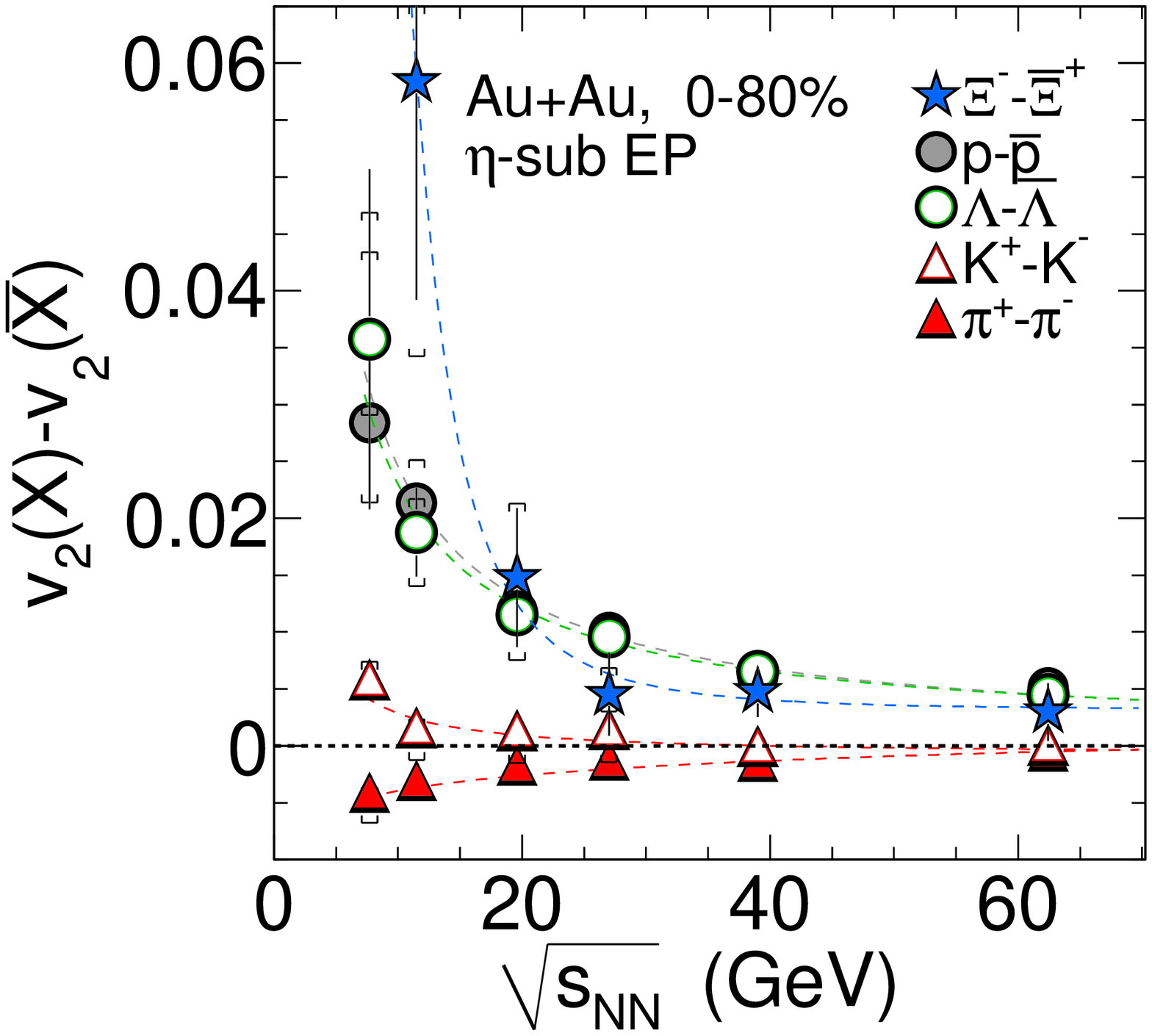}
\caption{
(Color online) 
The difference in $v_2$ between particles
and their corresponding anti-particles as a function of beam energy in
0--80\% Au+Au collisions.
The curves represent fits to data points as discussed in text.
Both statistical (vertical lines) and
systematic errors (caps) are shown. 
}
\label{fig_v2}
\end{minipage}
\end{figure}
Figure~\ref{fig_v1} shows the results from the beam energy scan. 
Plotted here is the $v_1$ slope ($dv_1/dy'$, where
$y'=y/y_{\rm{beam}}$ and $y$ is rapidity), near midrapidity as a function of beam energy for the mid-central
(10--40\%) Au+Au collisions~\cite{Pandit:2012mq}. The pion and anti-proton $v_1$ slopes show
negative values for all the beam energies studied. The proton $v_1$ slope
changes sign while going from 7.7 GeV to 11.5 GeV and 
then stays negative up to 200 GeV. However, the net-protons $v_1$ slope 
(obtained using $v_1$ slopes of $p$, $\bar{p}$ and ratio of $\bar{p}/p$)
changes sign from positive to negative and again becomes positive 
as a function of beam energy.
Both proton and net-proton $v_1$
slopes show a dip (or minimum) around $\sqrt{s_{NN}}$= 10--20 GeV. In order to
quantify the minimum position, it will be
interesting to add one more energy point around 15 GeV.
Also more theoretical as well as experimental studies are needed in order to understand these interesting observations.

\subsection{Elliptic Flow}
The elliptic flow $v_2$ is calculated as $\langle \cos2(\phi-\Psi_2)
\rangle$, where $\Psi_2$ is orientation of the second-order event plane.
Elliptic flow mainly probes the early stages of heavy-ion
collisions. 
At top RHIC energy of 200 GeV in Au+Au collisions, the elliptic flow scaled by the number
of constituent quarks ($n_q$) 
versus ($m_T-m_0)/n_q$ (where $m_T=\sqrt{p_T^2 + m_0^2}$) 
shows a scaling behavior where mesons and baryons have similar values
at intermediate $p_T$.
This is referred to as the number of constituent quark (NCQ) scaling~\cite{ncq_ref}. It is an established
signature of partonic matter formed in Au+Au collisions at 200 GeV and deviations from
such scaling would indicate the
dominance of hadronic interactions.
Hence
breaking of NCQ scaling at lower energies could be an indication of a ``turn-off''
of QGP signatures. 
\begin{figure}[htbp]
\begin{center}
\vspace{-0.4cm}
\includegraphics[width=0.8\textwidth]{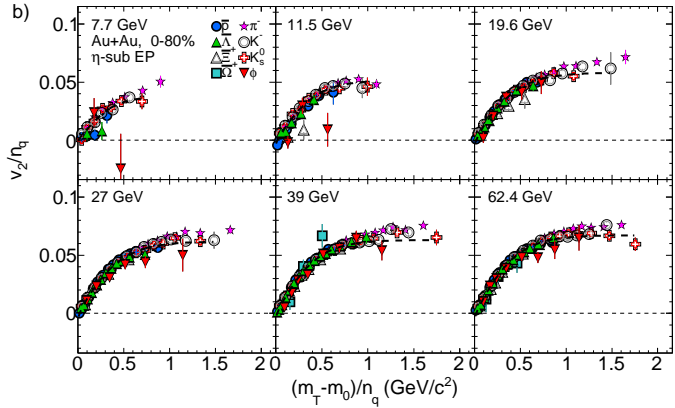}
\end{center}
\caption{(Color online)  $v_2/n_q$ as a function
  of $(m_T-m_0)/n_q$ for different particles in Au+Au collisions at $\sqrt{s_{NN}}$= 7.7, 11.5, 19.6, 27, 39 and 62.4 GeV.
  The errors shown are statistical only.
}
\label{fig_ncq_v2}
\end{figure}
Figure~\ref{fig_v2} shows the
difference in $v_2$ of particles and corresponding anti-particles as a
function of beam energy~\cite{v2_prl}. 
The curves represent fits to data points with functional form:
$f_{\Delta v_2}(\sqrt{s_{NN}})=a \sqrt{s_{NN}}^{-b}+c $.
The $v_2$ difference between particles and
anti-particles is observed to increase when we go towards the lower
energies. At low energies, $v_2(\pi^{-}) > v_2(\pi^{+}$), $v_2(K^{+}) >
v_2(K^{-})$, and $v_2(\rm{baryons}) > v_2$(anti-baryons). This
difference between particles and anti-particles suggests that the NCQ
scaling among particles and anti-particles is broken. However, the
observed difference between $v_2$ of particles and anti-particles could
be qualitatively explained by the models incorporating baryon
transport at midrapidity and hadronic
interactions~\cite{Dunlop:2011cf,Xu:2012gf}. 
We also observe that the baryons-mesons splitting for $v_2$
versus $m_T-m_0$  starts to disappear for
anti-particles at 11.5 GeV and below.
Figure~\ref{fig_ncq_v2} shows the $v_2/n_q$ versus $(m_T-m_0)/n_q$ for different particles
for $\sqrt{s_{NN}}$= 7.7--62.4 GeV~\cite{v2_prl}. We observe that results for all the particles
are consistent among each other within $\pm10$\% level, except for the
$\phi$-mesons at 7.7 and 11.5 GeV. At the largest $m_T-m_{0}$ the $\phi$-meson data points deviate
by 1.8$\sigma$ and 2.3$\sigma$ for $\sqrt{s_{NN}}=$ 7.7 and 11.5 GeV,
respectively. 
Since $\phi$-mesons have smaller hadronic interaction
cross-section, their smaller $v_2$ values could indicate that the
hadronic interactions start to dominate over partonic effects for the
systems formed at beam energies below $\sqrt{s_{NN}}=$ 11.5
GeV~\cite{phi_bed,phi2}. However, as can been seen from the figure, a
higher statistics data are needed to extend the $m_T-m_{0}$ range
and significance of the deviation observed. 

\subsection{Dynamical Charge Correlations}
The dynamical charge correlations are studied through a three-particle
mixed harmonics azimuthal correlator~\cite{3partcorr},
$\gamma=\langle
cos(\phi_\alpha+\phi_\beta-2\Psi_{\rm{RP}})\rangle$. This observable
represents the difference between azimuthal correlations projected 
onto the direction of the angular momentum vector and correlations 
projected onto the collision reaction plane.
\begin{figure}[htbp]
\begin{center}
\includegraphics[width=0.8\textwidth]{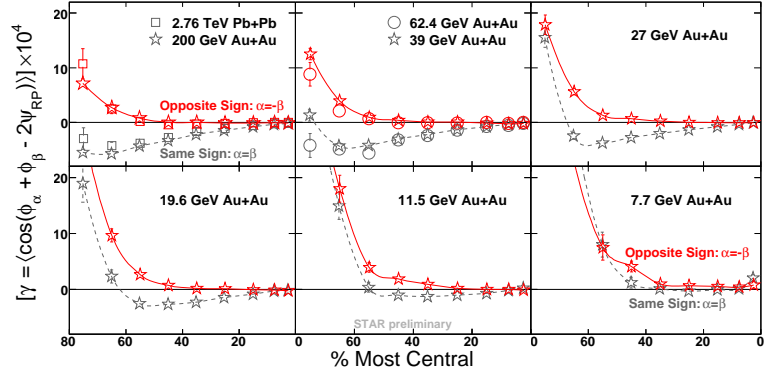}
\end{center}
\caption{(Color online)
Dynamical charge correlations 
as a function of centrality for Au+Au collisions 
at $\sqrt{s_{NN}}=$7.7-200 GeV. 
For comparison, results for Pb+Pb
collisions at 2.76 TeV are also shown. Errors are statistical only.}
\label{fig_dyn_ch}
\end{figure}
It is suggested that the difference in the correlations between same sign and
opposite sign charges in heavy-ion collisions could be related to local {\it parity}
violation if there is a deconfinement and a chiral phase
transition~\cite{Fukushima:2008xe}. This is also referred to as Chiral Magnetic Effect
(CME). At top RHIC energies, we observed a separation between the
correlations of same and opposite sign charges.
If this difference can be
attributed to the QCD phase transitions, the absence of such observation
could be an indication of the system which did not undergo the phase
transition. Hence, the observable could be useful to locate the energy in
the BES program where the QGP signature ``turns off''.
Figure~\ref{fig_dyn_ch} shows the results for the beam energies from
7.7--200 GeV as a function of centrality~\cite{gang}. For comparison, Pb+Pb
results from ALICE are also shown~\cite{Abelev:2012pa} 
which are observed to be consistent with the results from top RHIC energy. The
separation between same and opposite sign charges decreases with 
decreasing energy and vanishes below $\sqrt{s_{NN}}=$11.5 GeV.

\subsection{Nuclear Modification Factor}
Nuclear modification factor $R_{\rm{CP}}$ is one of the established
observable for the signature of QGP at top RHIC energy~\cite{rcp}. It is defined as ratio of yields
at central collisions to those at peripheral collisions, scaled by the
corresponding number of
binary collisions $N_{\rm{bin}}$. The number of binary collisions are
calculated from the Monte Carlo model.  It has been observed that at high $p_T$, the
$R_{\rm{CP}}$ of various particles is less than unity~\cite{rcp}, which is
attributed to the energy loss of the partons in the dense medium. In
the absence of dense medium, there may not be suppression of high $p_T$
particles, which can serve as an indication of ``turn-off'' of a QGP
signature. 
\begin{figure}[htbp]
\includegraphics[width=0.4\textwidth]{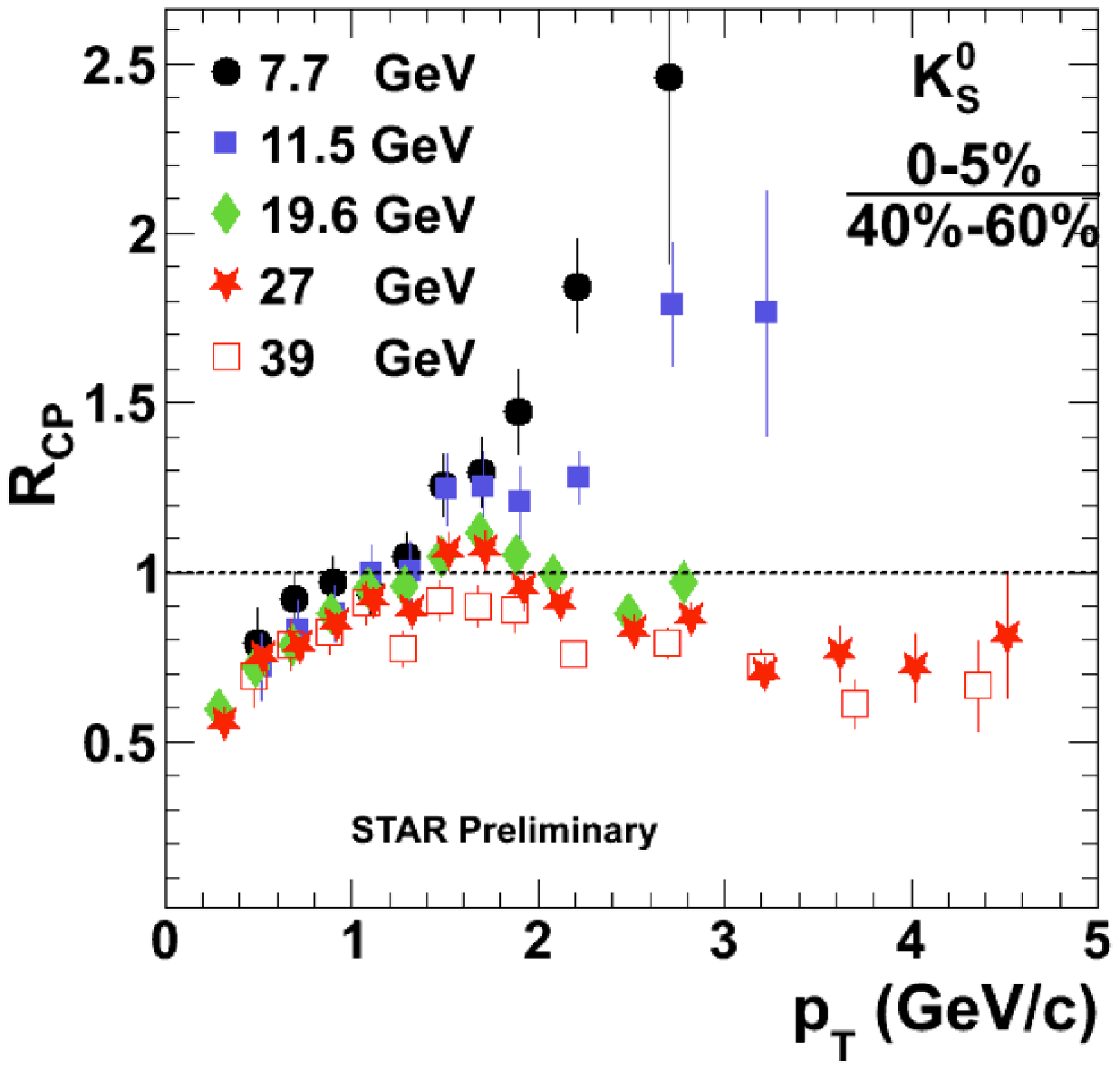}
\includegraphics[width=0.55\textwidth]{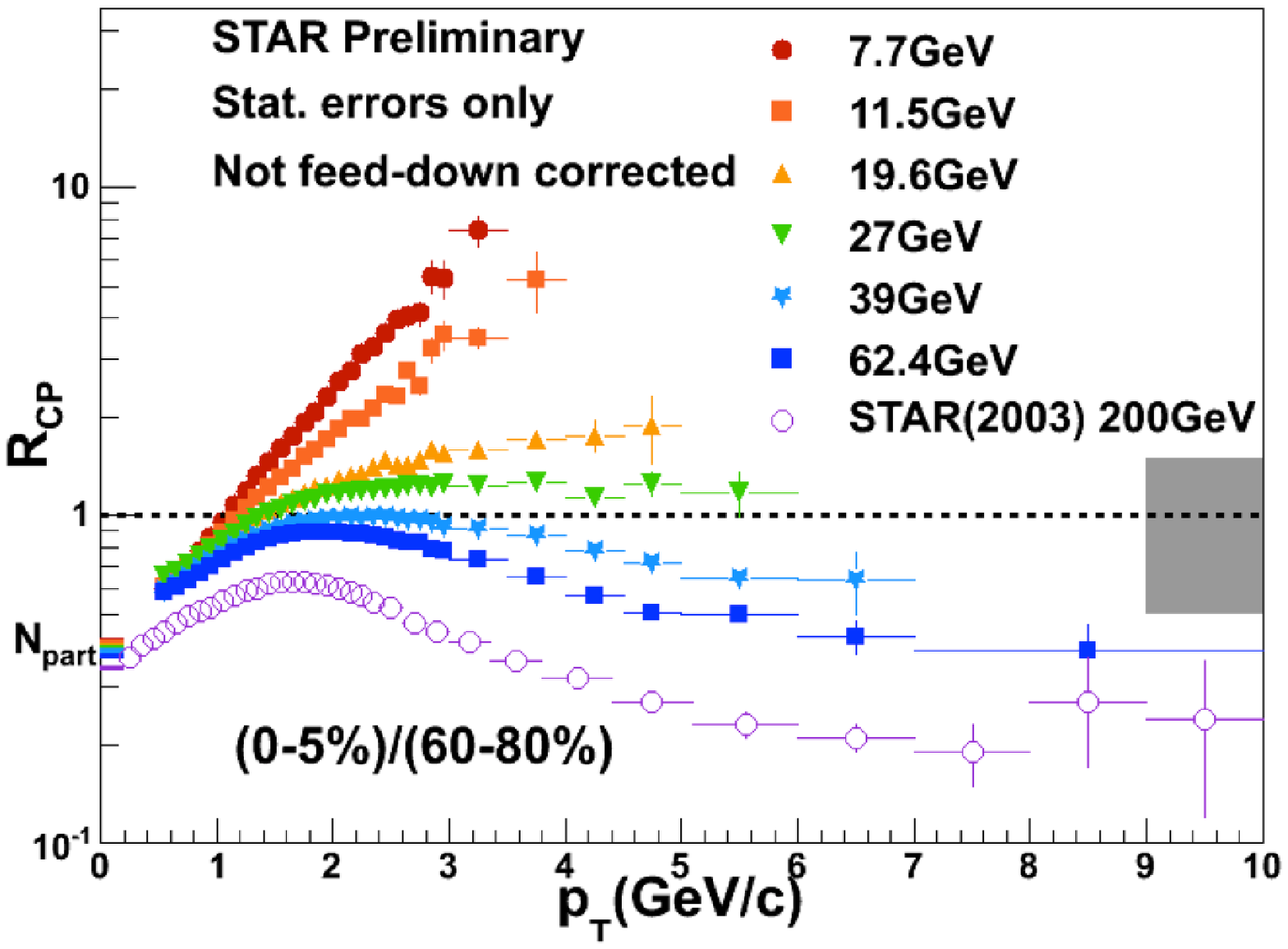}
\caption{(Color online) Left: $R_{\rm{CP}} \left( \frac{0-5\%}{40\%-60\%} \right)$ for
  $K^{0}_{S}$ in Au+Au collisions at  $\sqrt{s_{NN}}=$7.7--39 GeV.
Errors are statistical only. Right: $R_{\rm{CP}} \left( \frac{0-5\%}{60\%-80\%} \right)$ 
for charged hadrons in Au+Au collisions at  $\sqrt{s_{NN}}=$7.7--200 GeV.
Grey band corresponds to the systematic uncertainty.}
\label{fig_rcp}
\end{figure}

Figure~\ref{fig_rcp} (left panel) shows the $R_{\rm{CP}}$ of 
$K^{0}_{S}$
in Au+Au collisions at $\sqrt{s_{NN}}=$7.7--39 GeV~\cite{xiaoping}. We observe
that for $p_T>$ 2 GeV/$c$, the $R_{\rm{CP}} (K^0_S)$  is less than
unity at 39 GeV and then the value increases as the beam energy decreases. For
$\sqrt{s_{NN}}<$ 19.6 GeV, $R_{\rm{CP}} (K^0_S)$ is above unity,
indicating decreasing partonic effects at lower
energies. Figure~\ref{fig_rcp} (right panel) shows the $R_{\rm{CP}}$
results for charged hadrons in Au+Au collisions at
$\sqrt{s_{NN}}=$7.7--200 GeV~\cite{evan}. Again, we observe no suppression at lower
energies for $p_T>$ 2 GeV/$c$, supporting the $R_{\rm{CP}}
(K^{0}_{S})$ results. Both results suggest that partonic effects
become less important at lower energies and the cold nuclear matter
effects (Cronin effect) start to dominate at these energies~\cite{Cronin:1974zm}.

\begin{figure}[htbp]
\begin{minipage}{0.45\linewidth}
\centering
\vspace{-0.38cm}
\includegraphics[width=0.9\textwidth]{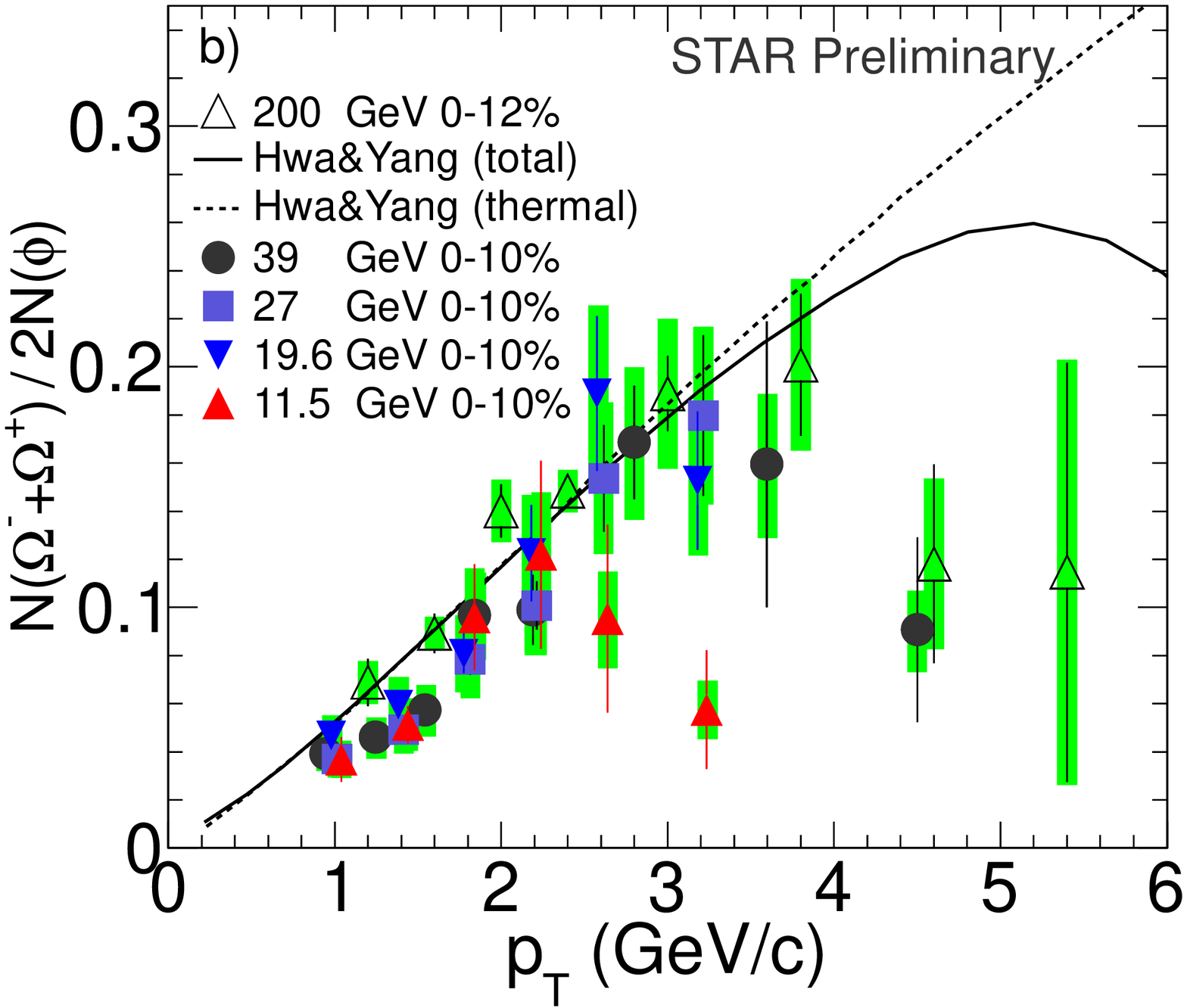}
\caption{(Color online) The baryon to meson ratio $N(\Omega^- +
  \Omega^+)/(2N\phi)$ as a function of $p_T$ in central Au+Au collisions
  at $\sqrt{s_{NN}}=$11.5--200 GeV. The curves represent model
  calculations by Hwa and Yang for $\sqrt{s_{NN}}=$ 200 GeV. Both statistical errors (vertical lines) and systematic
  errors (shaded bands) are shown.
}
\label{fig_omegaphi}
\end{minipage}
\hspace{0.4cm}
\begin{minipage}{0.45\linewidth}
\centering
\vspace{-1.0cm}
\includegraphics[width=1.08\textwidth]{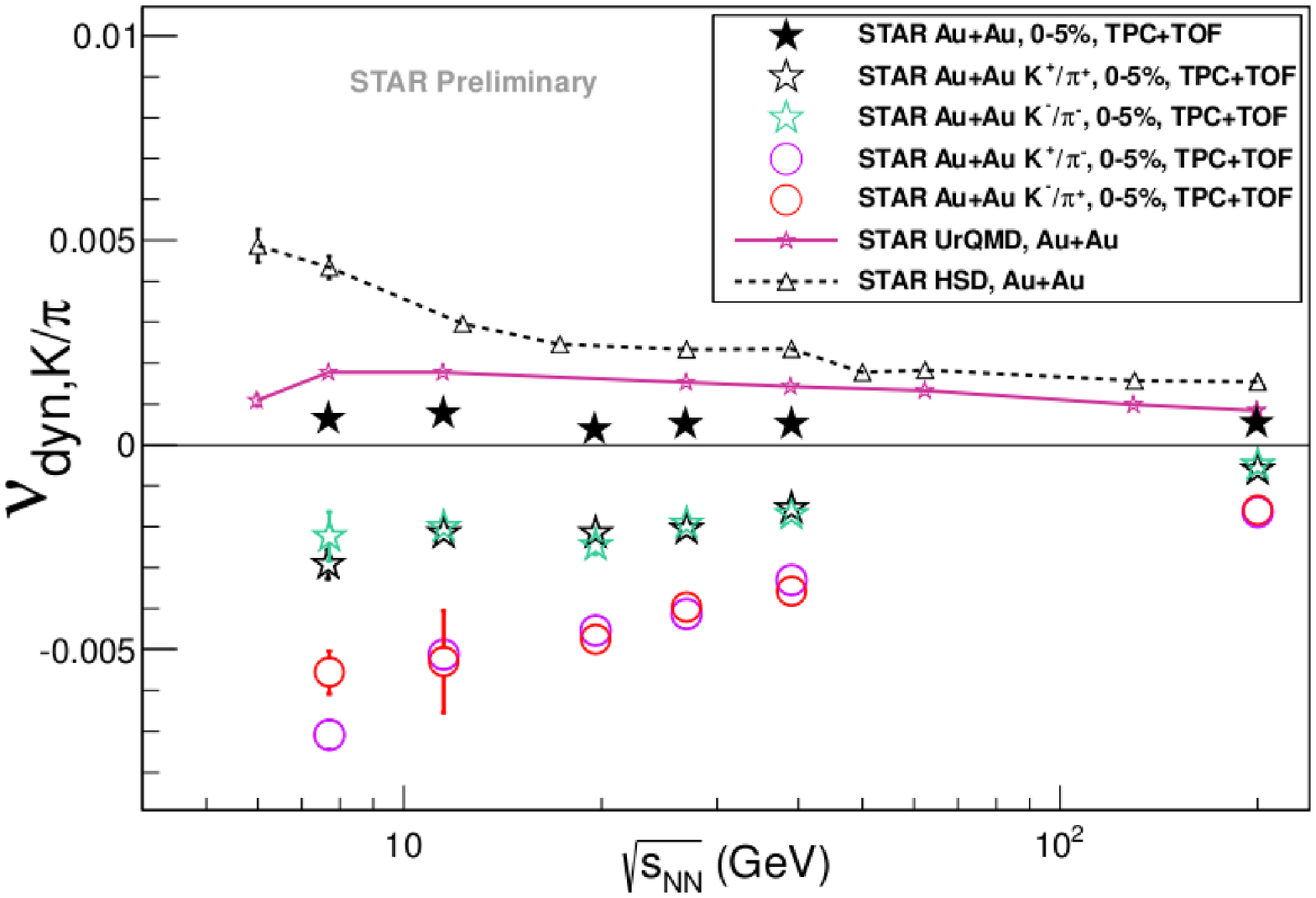}
\caption{
(Color online) $\nu_{\rm{dyn}}$ for $K/\pi$ (along with $K^+/\pi^+$, $K^-/\pi^-$ ,
$K^+/\pi^-$, and $K^-/\pi^+$) ratio in 0--5\% central Au+Au collisions are shown as a function
of energy. Results are compared with transport models such as UrQMD and
HSD. Errors are statistical.
}
\label{fig_k2pi}
\end{minipage}
\end{figure}
Figure~\ref{fig_omegaphi} shows the baryon to meson ratio $N(\Omega^- +
  \Omega^+)/(2N\phi)$ as a function of $p_T$ in central Au+Au collisions
  at $\sqrt{s_{NN}}=$11.5--200 GeV. The curves represent model
  calculations by Hwa and Yang in central collisions at $\sqrt{s_{NN}}=$ 200
  GeV~\cite{Hwa:2006vb,Hwa:2002zu} which assume the $\Omega$ and $\phi$ yields to
  be generated from the recombination of thermal strange quarks having
  exponential $p_T$ distribution. The particle ratio results at $\sqrt{s_{NN}}=$
  19.6, 27, and 39 GeV seem to follow that of 200 GeV, indicating a maximum
  around $p_T \ge$ 3 GeV/$c$ and then turning down as the $p_T$ is
  increased. 
  However, results at 11.5 GeV show different
  behavior i.e. show a maximum at somewhat lower $p_T$ of $\sim$2
  GeV/$c$ before turning down for higher values of $p_T$. This
  observation suggests that there might be a significant change in the
  underlying $p_T$ distributions of strange quarks recombining to form
  final $\Omega$ and $\phi$ for $\sqrt{s_{NN}} = $11.5 GeV and those
  for $\sqrt{s_{NN}} \ge$ 19.6 GeV.
 
\section{Search for QCD Critical Point}
In this section, we discuss the observables that could possibly be
related to the critical point search. First, we discuss the particle ratio
fluctuations such as $K/\pi$ ratio fluctuations as a function of beam
energy. After that, we discuss about the conserved number fluctuations
that include net-proton higher moments results. 

\subsection{$K/\pi$ Ratio Fluctuations}
If a system passes close to a critical point, large density variations
or enhanced fluctuations are expected e.g. as seen in critical opalescence. From experimental side, one
expects a non-monotonic variation of a potential
observable as a function of beam energy. Dynamical particle ratio
fluctuations such as $K/\pi$, $p/\pi$, and $K/p$, might be sensitive
to the initial state fluctuations arising from the existence of
critical point~\cite{nu_dyn1,nu_dyn2}. The observable used to quantify
these (e.g. $K/\pi$) dynamical
fluctuations $\nu_{\rm{dyn}}$ is given by
\begin{equation}
\nu_{dyn,K/\pi}=\frac{\langle N_K(N_K-1)\rangle}{\langle N_K\rangle^2} +
  \frac{\langle N_\pi(N_\pi-1)\rangle}{\langle N_\pi\rangle^2} - 2
    \frac{\langle N_K N_\pi\rangle}{\langle N_K\rangle \langle N_\pi\rangle }, 
\label{eqn_nudyn}
\end{equation}
where $N_K$ and  $N_\pi$ are the average number of kaons and pions in
an event, respectively. For a pure Poisson distribution,
$\nu_{dyn,K/\pi}$ will be zero. Figure~\ref{fig_k2pi} shows the
$\nu_{\rm{dyn}}$ results for $K/\pi$ (along with $K^+/\pi^+$, $K^-/\pi^-$ ,
$K^+/\pi^-$, and $K^-/\pi^+$) ratio in 0--5\% central collisions as a function
of beam energy\cite{tribedy}. The dynamical $K/\pi$ ratio fluctuations show a
smooth or monotonic behavior as a function of beam energy. The transport models such as HSD~\cite{Gorenstein:2008et} and
UrQMD~\cite{Bleicher:1999xi} show a similar smooth dependence on beam energy as observed in data.

\subsection{Conserved Number Fluctuations}
Higher moments of conserved number fluctuations are proposed to be 
potential observables for the search of the critical point~\cite{higher_mom1,Gupta:2011wh,hm3}.  
For a static, infinite medium, the correlation length $\xi$  diverges at
critical point. The various moments of event-by-event conserved
numbers (such as net-baryons, net-charge, and net-strangeness) distributions are
related to different powers of the correlation length. Higher moments
such as 
skewness $S$ and
kurtosis $\kappa$ are related to higher power of the correlation
length~\cite{corr_length,cl2}. Thus, these higher moments 
have a better sensitivity for the search of the
critical point. 
It has been proposed that the appropriate products of these moments such as
$\kappa\sigma^2$ and $S$$\sigma$ can be related to the ratios of order
susceptibilities calculated in lattice QCD and HRG
model as 
$\kappa\sigma^2=\chi^{(4)}_B/\chi^{(2)}_B$ and
$S\sigma=\chi^{(3)}_B/\chi^{(2)}_B$~\cite{chib1,Cheng:2008zh}. 
One of the advantages of using these
products or ratios is that they cancel the volume effects which are
difficult to estimate in an experiment. So in this
way, one can relate the experimental measurements with the lattice QCD
observables for the search of critical point. Since in an experiment,
it is difficult to obtain total baryons on an event-by-event basis,
net-protons are used as a proxy for the net-baryons. 
\begin{wrapfigure}{l}{0.58\textwidth}
   \includegraphics[width=0.55\textwidth]{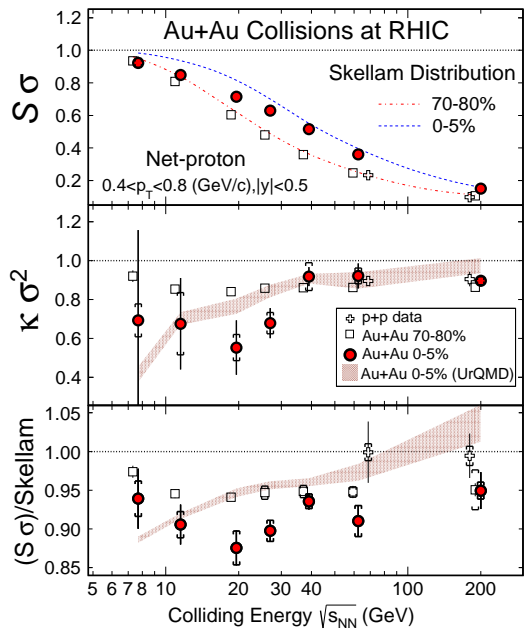}
 \caption{(Color online) $\kappa\sigma^2$, $S\sigma$ and $S\sigma$
values normalized by the Skellam expectations as a function of collision energy and two different centralities. Results from p + p collisions are also shown. All the results presented are corrected for detector efficiency. Results from UrQMD model calculations are also shown. The widths of bands represent statistical uncertainties. The error bars on data points are statistical while caps represent the systematic errors.}
\label{fig_cp}
\end{wrapfigure}
Figure~\ref{fig_cp} shows the $\kappa\sigma^2$ and $S\sigma$ for
net-protons as a function of beam energy for different collision
centralities~\cite{xiaofeng,netp_paper}. For comparison, the results
are shown for Skellam expectations and UrQMD model calculations that
do not include critical point~\cite{Bleicher:1999xi}. The results from
$p+p$ collisions at 200 GeV are also shown. The bottom panel shows the
$S\sigma$ values normalized by the corresponding Skellam
expectations. We observe that the moment products $\kappa\sigma^2$
and $S\sigma$ show similar values for central
for central 0--5\% and peripheral collisions (70--80\%)  for
$\sqrt{s_{NN}}=$ 39--200 GeV. For beam energies 
below 39 GeV, 
they have different values for central and peripheral
collisions. These values are below Skellam expectations for $\sqrt{s_{NN}}>$ 7.7 GeV for 0--5\% central
collisions. The deviation from Skellam expectation is observed to be more significant at
$\sqrt{s_{NN}}=$19.6 and 27 GeV. 
The UrQMD model calculations show a smooth monotonic behavior as a
function of collision energy. 
There are large
uncertainties for data points below 19.6 GeV that call for higher
statistics data at these energies.
In addition, a direct comparison to QCD calculations with
critical point obtained using similar dynamics at that of heavy-ion collisions
can provide definite answer about the existence of critical point. 

\section{BES Phase-II}
The first phase of the BES program has yielded several promising results for the
understanding of QCD phase diagram. Some of the observables require
high statistics data to make definite statements. These include
$\phi$-meson $v_2$ to test the NCQ scaling hypothesis at lower
energies and higher moments of net-protons to see whether there is a
non-monotonic variation towards lower energies that could suggest a
possible critical point. 
\begin{table}[h]
\tbl{Proposed energies, $\mu_B$ values, and required number of events for the BES
  Phase-II. Also listed are the corresponding fixed target
  $\sqrt{s_{NN}}$, centre of mass rapidity, and $\mu_B$ reach. 
}
{\begin{tabular}{c|c|c|c|c|c}
\hline
\multicolumn{3}{l|}{~~~~~~~~~~~~~~~~~~~BES Phase - II}
 &\multicolumn{3}{l}{~~~~~Fixed Target Collisions}\\  [0.1cm]
\cline{1-6} \\ [-0.3cm]
 $\sqrt{s_{NN}}$ (GeV) & $\mu_B$ (MeV) & $N_{\rm{event}}$ (Million)  &
 $\sqrt{s_{NN}}$ (GeV) & $y_{\rm{CM}}$  & $\mu_B$  (MeV) \\
\hline
 19.6   & 205 & 400   & 4.5  &1.52  &  585  \\
 15      & 250 &  100  & 4.0  &1.39  & 620\\ 
 11.5   & 315 & 120   & 3.5  &1.25  &  670 \\
  7.7    & 420 & 80     & 3.0  &1.05  & 720  \\
\hline
\end{tabular}\label{ta2} }
\end{table}
In addition, energy dependence of some
observables suggest to have a need of one more energy point around 15
GeV. For example, proton and net-proton $v_1$ slopes suggest a minimum
around 11.5--19.6 GeV as a function of energy which could 
be related to the softest point in equation of state. 
\begin{wrapfigure}{r}{0.6\textwidth}
    \includegraphics[width=0.7\textwidth]{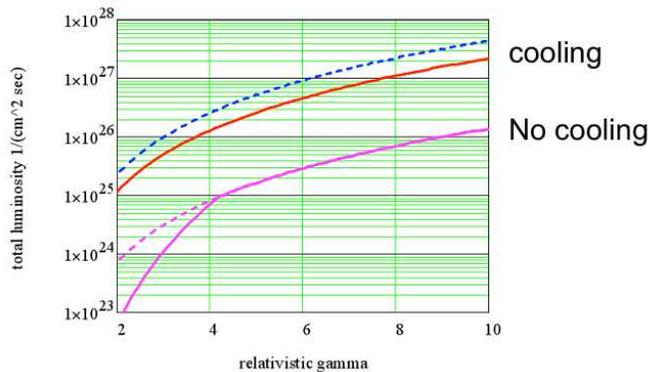}
  \caption{(Color online) Improvement in RHIC luminosity for the lower
  energies with electron cooling and long bunches (with space charge
  tune spread $\Delta Q_{\rm{SC}}=$
  0.05 and $\sigma_s=$ 3 m.)}
\label{fig_ecool}
\end{wrapfigure}
Having one more energy
point in between would indicate the exact location of the minimum. Similar reason
(although there is a monotonic variation as a function of beam energy
at the moment) might be argued for the freeze-out eccentricity. 
For net-proton higher moments, adding 15 GeV along with high-statistics data at
lower energies might provide the clear energy dependence trend with
high significance. 
One more energy point at 15 GeV is also important in view of the
fact that the gap between 11.5 and 19.6 GeV in terms of $\mu_B$ is more than 100 MeV.

For the reasons mentioned above, RHIC has decided to continue the exploration
of QCD phase diagram and hence proposed a second phase of the BES
program. The proposal for BES Phase-II includes high statistics data
below 20 GeV as listed in the Table~\ref{ta2}. 
To achieve the high statistics data at lower energies, an 
electron cooling device is requested to be installed at RHIC for increasing the
beam  luminosity~\cite{ecool}.
Simulation results (see Fig.~\ref{fig_ecool}) indicate
that with electron cooling, a significant improvement can be made to
increase luminosity (as shown by red solid curve in the figure). 
An additional improvement in luminosity (as shown by blue dashed
curve)  may be possible by operating with
longer bunches at the space-charge limit in the
collider~\cite{longbunch}. 
Electron cooling may increase the luminosity by a factor of 3--10 
and with longer bunches the luminosity may be increased by another factor of 2-5.
The high statistics data from BES Phase-II will not only allow the
precision measurements of the important observables discussed 
here but will also be helpful in the measurements of rare probes such as dilepton
production and hypertriton measurements at lower energies~\cite{rare1,rare2}. 

To maximize the use of collisions provided at STAR,
a fixed target proposal is made along with BES Phase-II.
The idea is to install a fixed Au target inside the STAR beam pipe to perform the Au(beam)-Au(target)
collisions. Such collisions will provide lower reach for the center of
mass energies and higher reach for the $\mu_B$ values for a given
BES Phase-II energy. The beam
energies and the $\mu_B$ values for the fixed target collisions are
listed in the Table~\ref{ta2} corresponding to the proposed BES
Phase-II energies. 
The  $\mu_B$ values are obtained from the parameterizations in
Ref.~\cite{Cleymans:2005xv}.
Clearly, it provides an opportunity to
reach the large values of $\mu_B$ and hence to explore the larger
portion of the QCD phase diagram. One of the advantages for such a proposal is that the
data taking for these fixed target collisions can be done concurrently during the normal RHIC
running and hence it will not affect the normal RHIC operations. 

These programs will also benefit from the proposed inner sector
upgrade of STAR TPC called the iTPC upgrade~\cite{itpc}.
At the moment,
inner sector of the TPC has the following issues: the inner sector
wires are showing the signs of ageing and unlike the outer TPC
sectors, it does not have the hermetic coverage at all radii. The
spacing between the rows is greater than 5 cm which results in missing
rows. To overcome these issues, it has been proposed to increase the
segmentation on the inner pad plane and renew the inner sector
wires. Simulation studies suggest that with iTPC upgrade it is
possible obtain better momentum resolution, better $dE/dx$ resolution
for particle identification,  and improved acceptance at higher
pseudorapidity $\eta$ and low $p_T$. At the moment, TPC $\eta$
coverage is about $|\eta|<$ 1.0, however, with iTPC upgrade, it might reach
$|\eta|<$ 1.7.
Similarly, lowest $p_T$ achieved can be as low as 60 MeV/$c$ compared to the present value of 125 MeV/$c$. 
 The above listed improvements will definitely
strengthen the technical aspects in the Physics analyses
proposed for the BES Phase-II. The BES Phase-II is expected to start
around 2018-2019.

\section{Summary}
The BES Phase-I enables RHIC to cover large range of $\mu_B$ (20--400 MeV)
in the phase diagram. At lower energies, a centrality dependence of freeze-out
parameters is observed. The observables such elliptic flow $v_2$,
nuclear modification factor $R_{\rm{CP}}$,
baryon to meson ratio $\Omega/\phi$, dynamical charge correlations, suggest that
hadronic interactions dominate for $\sqrt{s_{NN}} \le$ 11.5 GeV or
that the system did not undergo QGP phase transition at these lower
energies. The (net)-proton directed flow 
$v_1$ slope
show interesting behavior for the 
energy range $\sqrt{s_{NN}}<$ 20 GeV. 
The proton $v_1$ slope changes sign between
7.7 and 11.5 GeV. The net-proton $v_1$ slope changes sign twice
as a function of beam energy. 
Both proton and net- proton $v_1$ slopes
show a minimum around
11.5--19.6 GeV. The $\kappa \sigma^2$ and $S\sigma$ 
for net-protons show most significant deviations from Skellam
expectations at $\sqrt{s_{NN}} =$ 19.6 and 27 GeV.
BES Phase-II along with electron cooling, fixed
target proposal, and iTPC upgrade provides optimistic future for the
exploration of the QCD phase diagram, and hence for critical point and
phase boundary search.

We thank Prof. D. Keane, Prof. B. Mohanty, and Prof. Nu Xu for
reading the manuscript and providing helpful comments and suggestions.


\begin{thebibliography}{0}

\bibitem{qgp}  J. Adams {\it et al.} (STAR Collaboration), Nucl. Phys. A
  {\bf 757}, 28 (2005).

\bibitem{qgp_prop} V. Roy, A. K. Chaudhuri and B. Mohanty,
  Phys. Rev. C {\bf 86}, 014902 (2012).


\bibitem{ref_bes} M. M. Aggarwal {\it et al.} (STAR Collaboration), arXiv:1007.2613.

\bibitem{kumar_npa} L. Kumar (for STAR Collaboration), Nucl. Phys. A {\bf 862}, 125 (2011);

\bibitem{mohanty_npa} B. Mohanty, Nucl. Phys. A {\bf 830}, 899C (2009).




\bibitem{nsac} USA-NSAC 2007, Long range plan.

\bibitem{lattice1} Y. Aoki, G. Endroli, Z. Fodor, S. D. Katz and K. K.
  Szabo, Nature {\bf 443} 675, (2006).

\bibitem{ejiri} S. Ejiri, Phys. Rev. D {\bf 78}, 074507 (2008).
\bibitem{kapusta} E. S. Bowman and J. I. Kapusta, Phys. Rev. C {\bf 79}, 015202 (2009).

\bibitem{cp} M. A. Stephanov, Prog. Theor. Phys. Suppl. {\bf 153}, 139
  (2004) [Int. J. Mod. Phys. A {\bf 20}, 4387 (2005)] [hep-ph/0402115].

\bibitem{9gev} B. I. Abelev {\it et al.} (STAR Collaboration),
 Phys. Rev. C {\bf 81}, 024911 (2010).

\bibitem{lok} L. Kumar (for STAR Collaboration),
 J. Phys. G: Nucl. Part. Phys. {\bf 38}, 124145 (2011);
\bibitem{zhu} X. Zhu (for STAR Collaboration), Acta Phys. Polon. B Proc. Supp. {\bf 5}, 213 (2012).

\bibitem{stm} J. Adams {\it et al.} (STAR Collaboration), Nucl. Phys. A
 {\bf 757}, 102 (2005);


\bibitem{Andronic:2008ev} 
  A.~Andronic, F.~Beutler, P.~Braun-Munzinger, K.~Redlich and J.~Stachel,
  Phys.\ Lett.\ B {\bf 675}, 312 (2009)
  [arXiv:0804.4132 [hep-ph]].





\bibitem{Wheaton:2004qb} 
  S.~Wheaton and J.~Cleymans,
  Comput.\ Phys.\ Commun.\  {\bf 180}, 84 (2009)
  [hep-ph/0407174].




\bibitem{Kumar:2012fb} 
  L.~Kumar [STAR Collaboration],
  Nucl.\ Phys.\ A {\bf 904-905}, no. issue, 256c (2013)
  [arXiv:1211.1350 [nucl-ex]].


\bibitem{Das:2012yq} 
  S.~Das [STAR Collaboration],
  Nucl.\ Phys.\ A {\bf 904-905}, 891c (2013)
  [arXiv:1210.6099 [nucl-ex]].




\bibitem{Andronic:2009jd} 
  A.~Andronic, P.~Braun-Munzinger and J.~Stachel,
  Nucl.\ Phys.\ A {\bf 834}, 237C (2010)
  [arXiv:0911.4931 [nucl-th]].



\bibitem{Cleymans:2005xv} 
  J.~Cleymans, H.~Oeschler, K.~Redlich and S.~Wheaton,
  Phys.\ Rev.\ C {\bf 73}, 034905 (2006)
  [hep-ph/0511094].


\bibitem{Schnedermann:1993ws} 
  E.~Schnedermann, J.~Sollfrank and U.~W.~Heinz,
  Phys.\ Rev.\ C {\bf 48}, 2462 (1993)
  [nucl-th/9307020].


\bibitem{Mount:2010ey} 
  E.~Mount, G.~Graef, M.~Mitrovski, M.~Bleicher and M.~A.~Lisa,
  Phys.\ Rev.\ C {\bf 84}, 014908 (2011)
  [arXiv:1012.5941 [nucl-th]].


\bibitem{Bleicher:1999xi} 
  M.~Bleicher, E.~Zabrodin, C.~Spieles, S.~A.~Bass, C.~Ernst, S.~Soff, L.~Bravina and M.~Belkacem {\it et al.},
  J.\ Phys.\ G {\bf 25}, 1859 (1999)
  [hep-ph/9909407].




\bibitem{Lisa:2011na} 
  M.~A.~Lisa, E.~Frodermann, G.~Graef, M.~Mitrovski, E.~Mount, H.~Petersen and M.~Bleicher,
  New J.\ Phys.\  {\bf 13}, 065006 (2011)
  [arXiv:1104.5267 [nucl-th]].


\bibitem{Brachmann:1999xt} 
  J.~Brachmann, S.~Soff, A.~Dumitru, H.~Stoecker, J.~A.~Maruhn, W.~Greiner, L.~V.~Bravina and D.~H.~Rischke,
  Phys.\ Rev.\ C {\bf 61}, 024909 (2000)
  [nucl-th/9908010].


\bibitem{Csernai:1999nf} 
  L.~P.~Csernai and D.~Rohrich,
  Phys.\ Lett.\ B {\bf 458}, 454 (1999)
  [nucl-th/9908034].


\bibitem{Stoecker:2004qu} 
  H.~Stoecker,
  Nucl.\ Phys.\ A {\bf 750}, 121 (2005)
  [nucl-th/0406018].


\bibitem{Pandit:2012mq} 
  Y.~Pandit [STAR Collaboration],
  Nucl.\ Phys.\ A {\bf 904-905}, 357c (2013).
  [arXiv:1210.5315 [nucl-ex]].


\bibitem{ncq_ref} J. Adams {\it et al.} (STAR Collaboration),
  Phys. Rev. Lett. {\bf 95}, 122301 (2005).

\bibitem{v2_prl} L. Adamczyk {\it et al.} (STAR Collaboration),
  Phys. Rev. Lett. {\bf 110}, 0142301 (2013).

\bibitem{Dunlop:2011cf} 
  J.~C.~Dunlop, M.~A.~Lisa and P.~Sorensen,
  Phys.\ Rev.\ C {\bf 84}, 044914 (2011)
  [arXiv:1107.3078 [hep-ph]].


\bibitem{Xu:2012gf} 
  J.~Xu, L.~-W.~Chen, C.~M.~Ko and Z.~-W.~Lin,
  Phys.\ Rev.\ C {\bf 85}, 041901 (2012)
  [arXiv:1201.3391 [nucl-th]].



\bibitem{phi_bed} B. Mohanty and N. Xu, J. Phys. G {\bf 36}, 064022 (2009).


\bibitem{phi2} M. Nasim, B. Mohanty and N. Xu, Phys. Rev. C {\bf 87}, 014903 (2013).

\bibitem{3partcorr} B. I. Abelev {\it et al.} (STAR Collaboration),
  Phys. Rev. Lett. {\bf 103}, 251601 (2009).

\bibitem{Fukushima:2008xe} 
  K.~Fukushima, D.~E.~Kharzeev and H.~J.~Warringa,
  Phys.\ Rev.\ D {\bf 78}, 074033 (2008)
  [arXiv:0808.3382 [hep-ph]].

\bibitem{gang} G. Wang (for STAR Collaboration), Nucl.\ Phys.\ A {\bf 904-905}, 248c (2013).


\bibitem{Abelev:2012pa} 
  B.~Abelev {\it et al.}  [ALICE Collaboration],
  Phys.\ Rev.\ Lett.\  {\bf 110}, 012301 (2013)
  [arXiv:1207.0900 [nucl-ex]].


\bibitem{rcp} M. A. C. Lamont (for the STAR Collaboration),
  J. Phys. Conf. Ser. {\bf 50}, 192 (2006).

\bibitem{xiaoping} X. Zhang (for STAR Collaboration), Nucl.\ Phys.\ A {\bf 904-905}, 543c (2013). 

\bibitem{evan} E. Sangaline (for STAR Collaboration), Nucl.\ Phys.\ A {\bf 904-905}, 771c (2013). 

\bibitem{Cronin:1974zm}
  J.~W.~Cronin, H.~J.~Frisch, M.~J.~Shochet, J.~P.~Boymond, R.~Mermod, P.~A.~Piroue and R.~L.~Sumner,
  Phys.\ Rev.\ D {\bf 11} (1975) 3105.


\bibitem{Hwa:2006vb} 
  R.~C.~Hwa and C.~B.~Yang,
  Phys.\ Rev.\ C {\bf 75}, 054904 (2007)
  [nucl-th/0602024].

\bibitem{Hwa:2002zu} 
  R.~C.~Hwa and C.~B.~Yang,
  Phys.\ Rev.\ C {\bf 66}, 025205 (2002)
  [hep-ph/0204289].


\bibitem{nu_dyn1}
C. Pruneau, S. Gavin, and S. Voloshin, Phys. Rev. C
{\bf 66}, 044904 (2002). 

\bibitem{nu_dyn2}
J. Adams {\it et al.} (STAR Collaboration), Phys. Rev. C {\bf 68},
044905 (2003).


\bibitem{tribedy}
P. Tribedy (for STAR Collaboration), Nucl.\ Phys.\ A {\bf 904-905}, 463c (2013). 



\bibitem{Gorenstein:2008et} 
  M.~I.~Gorenstein, M.~Hauer, V.~P.~Konchakovski and E.~L.~Bratkovskaya,
  Phys.\ Rev.\ C {\bf 79}, 024907 (2009)
  [arXiv:0811.3089 [nucl-th]].









\bibitem{higher_mom1} 
M. M. Aggarwal et al. (STAR Collaboration), Phys. Rev. Lett. {\bf
  105}, 022302 (2010) [arXiv:1004.4959 [nucl-ex]].

\bibitem{Gupta:2011wh} 
  S.~Gupta, X.~Luo, B.~Mohanty, H.~G.~Ritter and N.~Xu,
  Science {\bf 332}, 1525 (2011)
  [arXiv:1105.3934 [hep-ph]].

\bibitem{hm3} F. Karsch and K. Redlich, Phys. Lett. B {\bf 695}, 136 (2011).



\bibitem{corr_length} M. A. Stephanov, Phys. Rev. Lett. {\bf 102}, 032301
  (2009); 

\bibitem{cl2}
  M. A. Stephanov, Phys. Rev. Lett. {\bf 107}, 052301 (2011).

\bibitem{chib1}
R. V. Gavai and S. Gupta, Phys. Lett. B {\bf 696}, 459 (2011) [arXiv:1001.3796 [hep-lat]].

\bibitem{Cheng:2008zh} 
  M.~Cheng, P.~Hendge, C.~Jung, F.~Karsch, O.~Kaczmarek, E.~Laermann, R.~D.~Mawhinney and C.~Miao {\it et al.},
  Phys.\ Rev.\ D {\bf 79}, 074505 (2009)
  [arXiv:0811.1006 [hep-lat]].


\bibitem{xiaofeng} X. Luo (for STAR Collaboration),  Nucl.\ Phys.\ A
  {\bf 904-905}, 911c (2013). 


\bibitem{netp_paper} L. Adamczyk {\it et al.} (STAR Collaboration), arXiv:1309.5681.


\bibitem{ecool} A. Fedotov, and W. Fischer, Private communications, 2012.

\bibitem{longbunch} A. Fedotov and M. Blaskiewicz, 
BNL CAD Tech Note: C-A/AP/449 (February 10, 2012).

\bibitem{rare1} 
B. Huang (for STAR Collaboration),  Nucl.\ Phys.\ A {\bf 904-905}, 565c (2013). 

\bibitem{rare2} 
Y. Zhu (for STAR Collaboration),  Nucl.\ Phys.\ A {\bf 904-905}, 551c (2013). 

\bibitem{itpc}
Y. Xu (for STAR Collaboration), poster ``Inner TPC Upgrade at STAR'', 2013 RHIC \& AGS Annual Users’ Meeting, BNL, USA.



\end{thebibliography}
\end{document}